\documentclass{article}
\usepackage{spconf}
\usepackage{amsmath, graphicx, multirow, textcomp, xcolor}
\usepackage{algorithm}
\usepackage{algpseudocode}

\title{S\lowercase{e}C\lowercase{o}ST: Sequential Co-Supervision for Large Scale Weakly Labeled Audio Event Detection}
\name{Anurag Kumar, Vamsi Krishna Ithapu}
\address{Facebook Reality Labs\\Redmond, WA USA\\ \texttt{anuragkr@fb.com, vamsi.ithapu@oculus.com}}
\begin{document}
\ninept
\maketitle
\begin{abstract}
Weakly supervised learning algorithms are critical for scaling audio event detection to several hundreds of sound categories. 
Such learning models should not only disambiguate sound events efficiently with minimal class-specific annotation but also be robust to label noise, which is more apparent with weak labels instead of strong annotations. 
In this work, we propose a new framework for designing learning models with weak supervision by bridging ideas from sequential learning and knowledge distillation.
We refer to the proposed methodology as SeCoST (\textit{pronounced Sequest}) --- Sequential Co-supervision for training generations of Students. 
SeCoST incrementally builds a cascade of student-teacher pairs via a novel knowledge transfer method. Our evaluations on Audioset (the largest weakly labeled dataset available) show that SeCoST achieves a mean average precision of \textbf{0.383} 
while outperforming prior state of the art by a considerable margin.
\end{abstract}
\section{Introduction} 
\label{sec:intro}
In the past decade, supervised learning has been extensively studied for audio event recognition and detection (AED) \cite{virtanen2018computational}. While several classical machine learning and deep learning methods have been developed for AED using strong labels \cite{tmmsurvey, virtanen2018computational, piczak2015environmental, mcloughlin2015robust}, much of the recent progress has focused on efficiently leveraging \emph{weakly labeled} data \cite{kumar2016audio}. In such a weak labeling paradigm, audio recordings are only tagged for the presence or absence of sound events; unlike the \emph{strong labeling} alternative where explicit time stamps of sound events are required. Hence, the annotation efforts are substantially lower thereby giving the ability to scale AED to large datasets, \emph{e.g} Audioset \cite{gemmeke2017audio}. It has also now become an important part of the annual DCASE challenge on sound events and scenes \footnote{http://dcase.community/challenge2019/index}. 

Several authors have shown promising results on weakly labeled AED \cite{kong2019weakly, chou2018learning, mcfee2018adaptive, kumar2017knowledge}. A significant fraction of these works uses deep convolutional neural networks in one form or other. Some are driven by attention mechanisms in neural networks \cite{bahdanau2014neural}, so as to efficiently characterize the temporal occurrences of events in the audio recordings \cite{chou2018learning, yu2018multi, wang2019comparison}. Other approaches have incorporated recurrent neural networks to model the temporal attributes of sound events \cite{wang2019comparison, adavanne2017sound}. 

However, large scale AED using weakly labeled data remains an open problem. 
When the timestamps of event occurrences are not provided, one cannot use explicit example clips of sounds for training. 
This clearly makes it harder to learn the necessary features and characteristics that disambiguate different sound events. 
Additionally, noise and the presence of other irrelevant sounds complicates the learning, in particular with long recordings. 
Lastly, one can observe that noisy labels are also of concern in weak label learning \cite{kumar2019learning, fonseca2019learning}. 

In this work, we address some of the above issues by presenting a novel learning framework for sounds. 
The proposal derives ideas from two distinct, but partly related, learning paradigms --- sequential learning \cite{huang2007convex} and knowledge sharing \cite{silver2013lifelong}. 
In the seminal work on {\it Sequence of Teaching Selves} \cite{minsky1994society}, 
the authors hypothesize that the human learning goes through different stages of development, where each such stage is ``guided'' by previous stages. 
This is, in principle, similar to lifelong learning where new knowledge is accumulated while retaining previous (learned) experiences \cite{silver2013lifelong, parisi2019continual}.
The central theme of these works is sequential learning, or, learning over time.
Alternatively, there is recent interest in knowledge distillation (KD) through \emph{teacher - student} frameworks \cite{hinton2015distilling, ba2014deep, bucilua2006model}, 
where the main motivation is model compression i.e., constructing a smaller, low-capacity, student model that emulates a high-capacity high-performance teacher. 
These student networks are optimized based on some carefully designed divergence measures \cite{hinton2015distilling, furlanello2018born}. 

We tackle weakly labeled learning by constructing a sequence of reasonably well trained neural networks (on the weak labels), 
where each network in the series is designed to be better than the previous one. 
However, unlike the classical KD where ``one-shot" distillation is done from a single teacher to a student with the goal of compressing the model, the student here aims to match the performance of the teacher {\it while} also correcting for teacher`s errors. 
This entails constructing a cascade of student-teacher pairs and allowing the student to learn from teacher`s mistakes over \emph{multiple generations}. 

We do this by controlling the amount of transferable knowledge between consecutive generations.   
This helps in correcting the implicit noise associated with weak labels, while distilling the necessary knowledge needed for generalization. We refer to the proposed framework as {\bf SeCoST} --- {\it Se}quential {\it Co}-Supervision for training generations of {\it S}tudents from {\it T}eachers. 

The rest of the presentation is as follows. In Section \ref{sec:baseCNN}, we first describe the baseline deep convolutional network that drives the work-flow of the overall framework. 
We then describe our proposed {\bf SeCoST} framework in Section \ref{sec:secost}, followed by experimental evaluations and discussion in Section \ref{sec:expt}. 
We conclude the paper in Section \ref{sec:conc}.

\begin{figure*}[t!]
\centering
\includegraphics[width=0.90\linewidth]{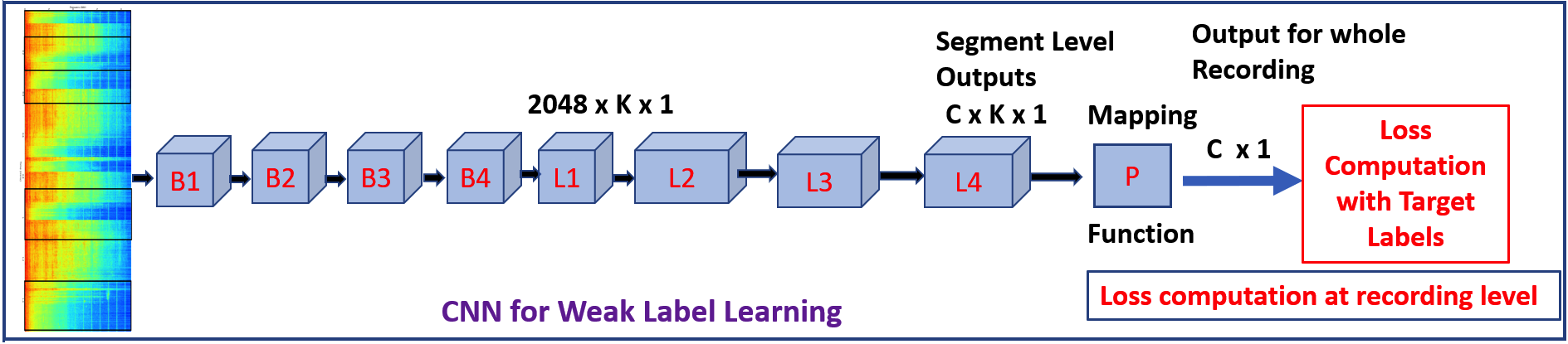}
\caption{\textbf{WELS-Net}: Deep CNN for weakly labeled AED. $|C|$: number of classes. $K$: number of segments obtained for the given input. $P$ maps the segment level output(s) to the recording level output.} 
\label{fig:netark}
\end{figure*}

\section{Deep CNN for Weakly Labeled  AED}
\label{sec:baseCNN}
\noindent {\bf Notation:} Let $\mathcal{S}$ be the set of audio recordings. $C$ denotes the set of labeled sound classes in these recordings. Each recording is represented by logmel features (denoted by $X$). 
Let $\mathbf{y} \in R^{|C|}$ be the (weak) label vector for the input $X$, and $\mathbf{y}_i = 1$ correponds to $i^{th}$ class being present (i.e., tagged).

We use a deep convolutional neural network as the prototypical learning model that drives our proposed framework \cite{kumar2017knowledge,shah2018closer}. Given an input audio recording, the idea is to first produce segment level predictions. The segments are audio snippets of small duration (e.g., $1$ second length). The resulting segment level outputs are then mapped to a recording level prediction. The appropriate loss is then calculated using this prediction and the recording level weak label. The mapping from segments to recording may be done via simple mean or max operation over segment level outputs or even by a neural network, if necessary. 


Figure \ref{fig:netark} and Table \ref{tab:cnnarch} summarize the network schematic and the specific architectural details. 
$B1$ to $B4$ blocks consist of two convolutional layers followed by a pooling layer. 
The size of segment level output at layer $L4$ will be $|C| \times K \times 1$, where $|C|$ is the number of classes and $K$ denotes the number of possible segments for a given input. The recording level output for any given class is obtained by taking the average of segment level outputs (layer P). 

We will refer to the above network as WELS-Net -- {\it WE}akly {\it L}abeled {\it S}ound {\it Net}work.  WELS-Net is flexible with respect to input size, thereby allowing processing of audio recordings of variable lengths. 
Moreover, it produces segment level outputs which can then be used for temporal localization of sound events in the audio recordings. 

Note that, WELS-Net takes Logmel spectrograms as input. Specifically, we use $64$ mel-filters and mel features are extracted using a $16$ ms window moving by a stride of $10$ ms. The sampling rate of all audio recordings is $16$ kHz. 
Table \ref{tab:cnnarch} shows output sizes at each layer for a logmel input with $1024$ frames (size $1024 \times 64$). 
For this input, we get $K=30$ segments at layer $L4$. In the time domain, this network produces outputs for \texttildelow $1$ second segments (with a stride of \texttildelow $0.33$ seconds). 

\begin{table}[th]
  \centering
  \caption{CNN Architecture Used. $F$: Number of Filters. $P$: Padding Size. $S$: Stride. $BN$: Batch Normalization. $|C|$: Number of classes. Last column shows output sizes for an input with $1024$ logmel frames, i.e $1024 \times 64$ dimensional input. }  
    \resizebox{1.0\columnwidth}{!}{
    \begin{tabular}{|c|c|c|}
    \hline
    \textbf{Layers} & \textbf{Layer Parameters}  & Output size    \\
    \hline 
    Input & - & $1 \times 1024 \times 64$ \\
    \hline
    \multirow{3}{*}{Block B1}   & $3 \times 3$ conv, F: 64, P:1, S:1 $\rightarrow$ BN $\rightarrow$ ReLU & $64 \times 1024 \times 64$ \\
                                               &  $3 \times 3$ conv, F: 64, P:1, S:1 $\rightarrow$ BN $\rightarrow$ ReLU  & $64 \times 1024 \times 64$\\
                                                & $4 \times 4$ pool,  S:4  & $64 \times 256 \times 16$ \\
    \hline
	\multirow{3}{*}{Block B2}   & $3 \times 3$ conv, F: 128, P:1, S:1 $\rightarrow$ BN $\rightarrow$ ReLU & $128 \times 256 \times 16$ \\
                                               &  $3 \times 3$ conv, F: 128, P:1, S:1 $\rightarrow$ BN $\rightarrow$ ReLU & $128 \times 256 \times 16$ \\
                                                & $2 \times 2$ pool,  S:2  & $128 \times 128 \times 8$ \\ 
    \hline
	\multirow{3}{*}{Block B3}   & $3 \times 3$ conv, F: 256, P:1, S:1 $\rightarrow$ BN $\rightarrow$ ReLU  & $256 \times 128 \times 8$ \\
                                               &  $3 \times 3$ conv, F: 256, P:1, S:1 $\rightarrow$ BN $\rightarrow$ ReLU & $256 \times 128 \times 8$ \\
                                                & $2 \times 2$ pool,  S:2  & $256 \times 64 \times 4 $ \\       
    \hline
    \multirow{3}{*}{Block B4}   & $3 \times 3$ conv, F: 512, P:1, S:1 $\rightarrow$ BN $\rightarrow$ ReLU & $512\times 64 \times 4 $ \\
                                               &  $3 \times 3$ conv, F: 512, P:1, S:1 $\rightarrow$ BN $\rightarrow$ ReLU & $512 \times 64 \times 4 $ \\
                                                & $2 \times 2$ pool,  S:2  & $512 \times 32 \times 2 $ \\ 
    \hline
    Layer L1   & $3 \times 2$ conv, F: 2048, P:0, S:1 $\rightarrow$ BN $\rightarrow$ ReLU & $2048 \times 30 \times 1 $ \\
	\hline 
	Layer L2   & $1 \times 1$ conv, F: 1024, P:0, S:1 $\rightarrow$ BN $\rightarrow$ ReLU & $1024 \times 30 \times 1 $ \\
	\hline
	Layer L3   & $1 \times 1$ conv, F: 1024, P:0, S:1 $\rightarrow$ BN $\rightarrow$ ReLU & $1024 \times 30 \times 1 $ \\
	\hline
	Layer L4   & $1 \times 1$ conv, F: C, P:0, S:1 $\rightarrow$ BN $\rightarrow$ ReLU & $|C| \times 30 \times 1 $ \\
	\hline
	P 	& Global Average Pooling & $|C|  \times 1 $  \\
	\hline      
    \end{tabular}}
  \label{tab:cnnarch}%
\end{table}%

\subsection{Network Training}
\label{sec:nettrain}

Recall that the inputs are logmel features denoted by $X$. We assume a realistic scenario where $X$ may be tagged with multiple labels, i.e., multiple sound events may be present in a single recording. The goal is to train a neural network $\mathcal{N(\theta)}$ which can generalize well on unseen data. 
The network is trained by minimizing a loss function which measures divergence between the network outputs $\mathcal{N}(\theta, X)$ and the target $\mathbf{y}$. Let $\mathcal{L}(\mathcal{N}(\theta,X),\mathbf{y})$ denote the loss function. 
We use binary-cross entropy loss. Thereby, we have
\begin{equation} \label{eq:loss}
\begin{aligned}
    l(p_i,y_i) = & - y_i\, \log(p_i) - (1- y_i)\, \log(1 - p_i) \\   
    \mathcal{L}(\mathcal{N}(\theta,X),\mathbf{y})  = & \frac{1}{|C|} \sum_{i=1}^{|C|} l(y_i,p_i)
\end{aligned}. 
\end{equation}
$p_i$ is the output of the network for the $i^{th}$ class. $l(y_i,p_i)$ is the loss for this $i^{th}$ class, and $\mathcal{L}(\mathcal{N(\theta,X)},\mathbf{y})$ computes the overall loss for the input $X$ and the corresponding target $\mathbf{y}$.  

\section{SeCoST framework}
\label{sec:secost}

Using the network architecture presented in Section \ref{sec:baseCNN}, we will now describe the proposed sequential co-supervision learning. 
As mentioned in Section \ref{sec:intro}, SeCoST follows the principle of sequence of teaching selves. A sequence of learners (neural networks here) are trained. At each stage, the learning of a new network is supervised by the already trained network(s) from previous stages.
First recall that training a network with the loss function in Eq \ref{eq:loss} corresponds to learning from available ground truth labels $\mathbf{y}$.
This is our initialization a.k.a. the first teacher. We denote this {\it base model} by $\mathcal{N}^{T_0}$. 
Once it is learned, we propose that it can co-supervise training a new network (identical in architecture to the teacher) from scratch. 
In other words, the outputs of the teacher networks drive the supervision in future generations. We now formalize this process. 

If the vector $\hat{\mathbf{y}}$ denotes the output of the teacher with input $X$, 
then the new target denoted by $\bar{\mathbf{y}}$, for the same input $X$, is given by
\begin{equation} \label{eq:mod}
\bar{\mathbf{y}} = \alpha \mathbf{y} + (1 - \alpha) \hat{\mathbf{y}}
\end{equation}
where $\alpha$ is a hyper-parameter that controls the contribution of the teacher network`s supervision. 

\setlength{\textfloatsep}{1pt}
\begin{algorithm}[t!]
   \caption{\textbf{SeCoST}:}
       \textbf{Input}: Training data $\mathcal{D}=\{X^i,\mathbf{y}^i$\}, Number of stages $S$, \{$\alpha^{s}$ for each stage $s$ = 1 to $S$ \} \\
    \textbf{Output}: Trained Network after $S$ stages
    \begin{algorithmic}[1]
    \State Train base WELS-Net or Teacher-0 ($\mathcal{N}^{T_0}$) using $\mathcal{D}=\{X^i,\mathbf{y}^i$\}
    \For{$s$ = $1,2,....S$}
        \State Compute new target $\bar{\mathbf{y}}^i$ for all training points ($X^i,\mathbf{y}^i$) from $\mathbf{y}^i$, $\alpha^s$ and prediction of $\mathcal{N}^{T_{s-1}}$ on $X^i$ using Eq \ref{eq:mod}
        \State Train new WELS-Net ($\mathcal{N}^{S_{s}}$)  for current stage using $\mathcal{D} = \{X^i,\bar{\mathbf{y}}^i\}$
        \State $\mathcal{N}^{T_{s}}$ = $\mathcal{N}^{S_{s}}$ // Student becomes teacher for next stage
    \EndFor
    \State return $\mathcal{N}^{T_{S}}$
\end{algorithmic}
\label{alg:secost}
\end{algorithm}


With a single teacher, the class-wise target for the student model is $\bar{y}_i = \alpha y_i + (1-\alpha) \hat{y}_i$. The corresponding loss for the $i^{th}$ class now becomes

\begin{flalign} 
\label{eq:modcl}
& l(p_i,\bar{y}_i)  = - \bar{y}_i\, \log(p_i) - (1- \bar{y}_i)\, \log(1 - p_i) \nonumber \\ 
& = -\alpha y_i  p_i - (1-\alpha y_i)\log(1-p_i) \,\, + (1-\alpha)\, \hat{y}_i \, \log\frac{1-p_i}{p_i} \nonumber \\ 
& l(p_i,\bar{y}_i) = l(p_i,\alpha y_i) + (1-\alpha) \hat{y}_i \, \log\frac{1-p_i}{p_i}
\end{flalign}

Using this new class-wise loss, the overall new loss function for the student network, denoted by $\mathcal{L}(;,\bar{\mathbf{y}})$, is
\begin{align} 
\label{eq:newloss}
\mathcal{L}(;,\bar{\mathbf{y}}) & =  \frac{1}{|C|} \sum_{i=1}^{|C|} [l(\alpha y_i,p_i) + (1-\alpha) \hat{y}_i \log\frac{1-p_i}{p_i}] \nonumber\\
& =  \mathcal{L}(;,\alpha \mathbf{y}) +  (1-\alpha) \frac{1}{|C|} \sum_{i=1}^{|C|}  \hat{y}_i \log\frac{1-p_i}{p_i}
\end{align}
which is a combination of loss w.r.t. ground truth $\mathbf{y}$, although, weighted by a factor of $\alpha<1$; and a term representing supervision from the teacher. We can rewrite this as 
\begin{equation} \label{eq:newloss1}
\mathcal{L}(;,\bar{\mathbf{y}}) = \mathcal{L}(;,\mathbf{y}) +  (1-\alpha) \frac{1}{|C|} \sum_{i=1}^{|C|}  (y_i - \hat{y}_i) \log\frac{1-p_i}{p_i}
\end{equation}

Here, the first term is same as computing the loss w.r.t. $\mathbf{y}$. The second term now involves both the ground truth labels and the teacher's predictions. Hence, the additional supervision for student is entirely determined by how much the teacher's predictions ($\hat{y}_i$) differ from the true labels $y_i$, i.e., the student is trying to learn from the mistakes made by the teacher network. 
This additional information is hypothesized to improve the generalization capabilities. 

This overall procedure can now be emulated to multiple stages $S$. Algorithm \ref{alg:secost} summarizes this sequential procedure over $S$ stages. The output of this procedure is a network trained over $S$ generations with one or more teachers in each stage. 
Note that we expect the improvement in generalization to have diminishing returns after some stages, and we discuss more about this behaviour in Section \ref{sec:expt}.

\noindent \textbf{Multiple Teachers per Stage:}
Observe that Eqs \ref{eq:mod} and \ref{eq:newloss1} are parameterizing learning from a single teacher at a given stage. 
The above procedure can also be extended to incorporate multiple teachers (denoted by $T$) per stage. 
The new target is given by a convex combination of all available supervision, from all the $T$ teachers, as shown below. 
\begin{equation} \label{eq:convcomb}
\bar{\mathbf{y}} =  \sum_{k=0}^T   \alpha_{k} \hat{\mathbf{y}}^k \,\,\, \mbox{and}\,\,\, \sum_{k=0}^T \alpha_k = 1 
\end{equation}
where $\hat{\mathbf{y}}^0 = \mathbf{y}$ represents the ground truth labels, 
and $\alpha_k$, $k=0 \,\,\mbox{to}\,\,T$, parameterize the contribution of the ground truth and the $T$ teacher networks respectively. The class-wise loss $l(p_i,\bar{y}_i)$ becomes
\begin{equation} \label{eq:modclmulti}
l(p_i,\bar{y}_i)  = l(p_i,\alpha_0 y_i) + \sum_{k=1}^T \alpha_k \hat{y}_i^k \log\frac{1-p_i}{p_i}
\end{equation}

\section{Experiments and Results}
\label{sec:expt}

We evaluate SeCoST using Audioset (the largest available dataset for sound events) \cite{gemmeke2017audio}. It has weakly labeled examples for $527$ sound events, with approximately $2$ million \emph{training} examples and $20$k \emph{evaluation} recordings. We use $25$k samples from the training set for validation. Each recording is $10$ seconds long (and we resample them at $16kHz$). Audioset is multi-label in nature with $\sim2.7$ labels per recording. 
PyTorch is used for implementing the networks \cite{paszke2017automatic}. Training utilizes Adam  \cite{kingma2014adam} where hyperparameters (like learning rates) are tuned using the validation set. Similar to existing AED works \cite{gemmeke2017audio,kumar2017knowledge,kong2019weakly}, 
Average precision (AP) and area under ROC (AUC) are used to measure the performance \cite{fawcett2004roc,buckley2004retrieval}. 
Further, mean average precision (mAP) and mean AUC (mAUC) over all classes summarizes the overall performances. In this work, we use a single teacher at each stage (Alg. \ref{alg:secost}), leaving evaluation of multiple teacher per stage for future work. 

\begin{table}[t!]
  \centering
  \caption{Comparing SeCoST with state of the art methods on Audioset.}  
    \resizebox{1.0\columnwidth}{!}{
    \begin{tabular}{|c|c|c|c|c|c|}
    \hline
	Method & mAP & mAUC & Method & mAP & mAUC \\
    \hline
        \cite{kong2019weakly} - Pooling   & 0.343 & 0.966 & \cite{kong2019weakly} - Attention & 0.361 & 0.969 \\
		\hline 
		\cite{kong2019weakly} - Attention-Large   & 0.369 &  0.969 &  TALNet \cite{wang2019comparison} - Pooling & 0.361   & 0.966 \\
		\hline
		TALNet\cite{wang2019comparison} - Attention & 0.362  & 0.965 &  WELS-Net (Our's Base) & 0.352 & 0.962 \\
		\hline
		\textbf{SeCoST (fixed $\alpha = 0.3$)} & \textbf{0.379} & \textbf{0.970} & \textbf{SeCoST (Variable $\alpha$)}  & \textbf{0.383}  & \textbf{0.971} \\
		\hline
    \end{tabular}}
  \label{tab:sotacomp}%
\end{table}%
 
\noindent \textbf{Performance Comparison}: 
Table \ref{tab:sotacomp} compares {\bf SeCoST} with existing state-of-the-art methods on Audioset.
Note that the authors of \cite{kong2019weakly} use embeddings from a network trained on a very large database of audio recordings (YouTube-70M) \cite{hershey2017cnn}, thereby the resulting feature representations already lead to enhanced performance. However, we work \emph{directly} with audio recordings and use their logmel representations. 
Our base WELS-Net model ($\mathcal{N}^{T_0}$) trained on the ground truth labels gives an mAP of $0.352$ over all the $527$ events. SeCoST gives an mAP of $\textbf{0.383}$, improving WELS-Net by $\mathbf{8.8\%}$. Notably, it is also $\mathbf{3.8\%}$ better than the best reported performance in literature. 
Our best performance of $0.383$ mAP is obtained by increasing the contribution of teacher ($1 - \alpha$) as the sequence progresses. Using a fixed $\alpha$ of $0.3$, we improve the base WELS-Net by $7.7\%$ (from $0.352$ to $0.379$). 

\begin{figure}[t]
\centering
\includegraphics[width=0.49\linewidth]{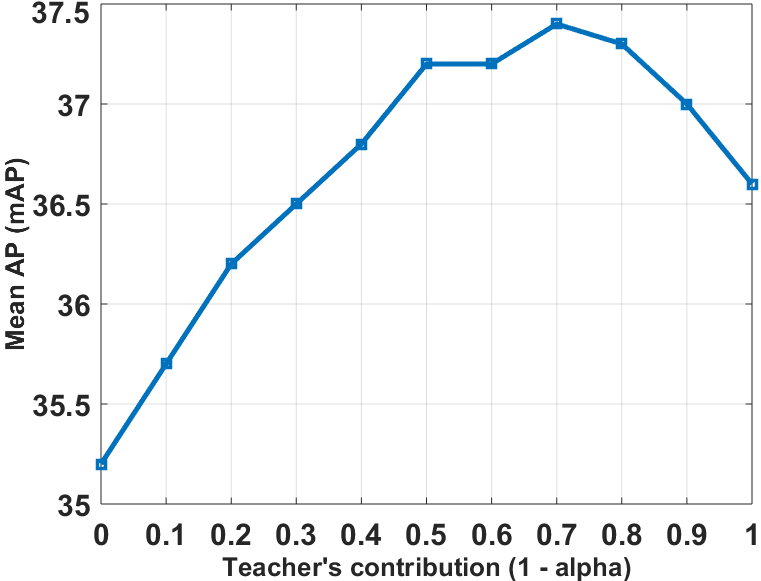}
\includegraphics[width=0.49\linewidth]{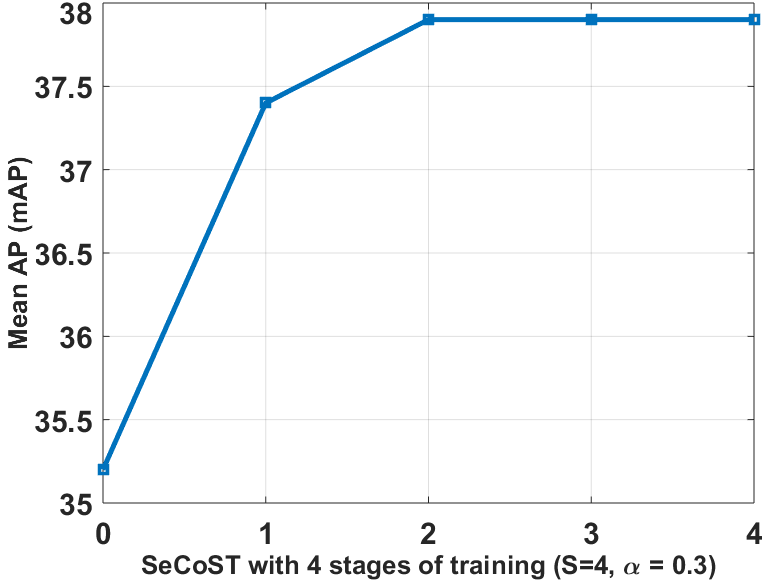}
\caption{\textbf{Left}: Single stage SeCoST with varying contribution ($1 - \alpha$) from teacher. Base WELS-Net ($\mathcal{N}^{T_0}$) as teacher. \textbf{Right}: $4$ stages of SeCost showing performance after each stage. Teacher's contribution fixed at $0.7$.}
\label{fig:alpha}
\end{figure}

\begin{figure}[t]
\centering
\includegraphics[width=0.49\linewidth]{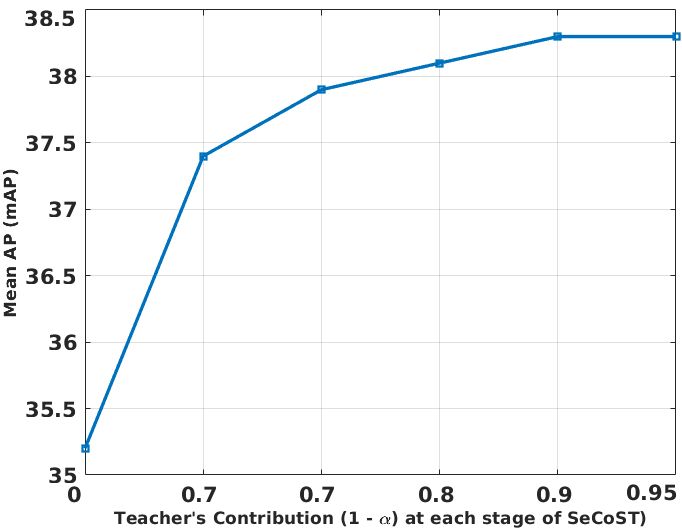}
\includegraphics[width=0.49\linewidth]{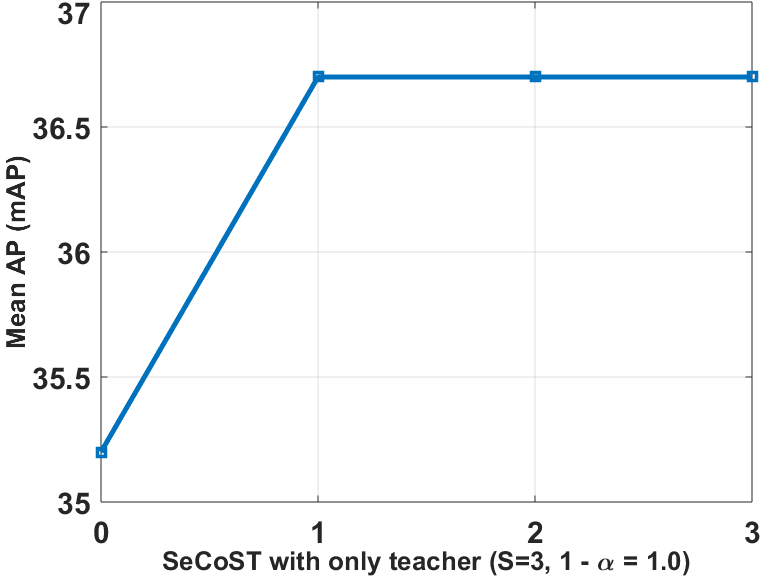}
\caption{\textbf{Left}: $5$ stage SeCoST with variable teacher's contribution. $x$-axis shows teacher's contribution at each stage. \textbf{Right}: SeCoST with only teacher networks ($\alpha = 0$)}
\label{fig:alpha_var}
\end{figure}

\noindent \textbf{Effect of } $\mathbf{\alpha}$: 
Figure \ref{fig:alpha} (Left) analyzes the influence of teacher's contribution (parameterized by $1 - \alpha$) in SeCoST (see Alg \ref{alg:secost}). Here a single stage of SeCoST is done with base WELS-Net as the teacher. $1 - \alpha = 0$ represents no contribution from the teacher, i.e training only on ground truth labels. We can see that the performance improves as teacher's contribution increases, but only up to a certain point. This occurs at $1 - \alpha = 0.3$, with the corresponding mAP of $0.374$. Thereby, improving the base WELS-Net by $6.3\%$. From this, we can argue that teacher`s supervision should be well proportioned along with ground truth supervision.

\noindent \textbf{SeCoST Stage-wise performance}: 
Fig \ref{fig:alpha} (Right) evaluates the sequential training aspect of SeCoST. We do $4$ stages of training in SeCoST (S = $4$ in Alg \ref{alg:secost}). The contribution of teacher in the supervision remains same for all 4 stages with $1 - \alpha_s = 0.7,\,\, for\,\,s = 1\,\,to\,\,4$. We can see that sequential co-supervision leads to improvements, although with diminishing returns after each stage. After the first stage of co-supervision ($\mathcal{N}^{T_0}$ as teacher), the mAP improves from $35.2$ to $37.4$, a $6.3\%$ improvement over the base WELS-Net. Using this improved network as teacher in the second stage of SeCoST, we see a further improvement of $1.3\%$ (overall a $7.6\%$ improvement from WELS-Net). The performance then saturates and we do not see any additional improvements in future stages.

We saw above that, SeCoST in general works better with lower $\alpha$ i.e., larger weight to teacher's contributions. This suggests that as newer generations of networks become teachers, it might be helpful to increase their contribution in co-supervision. To evaluate this, we run SeCoST for $5$ stages with $\{\alpha_1 = 0.3, \alpha_2 = 0.3, \alpha_3 = 0.2, \alpha_4 = 0.1, \alpha_5 = 0.05\}$. The performance at different stages is shown in Figure \ref{fig:alpha_var}. As teacher's contribution is increased from $0.7$ to $0.8$ at stage $3$, we see an improvement in performance, unlike the case where it was fixed at $0.7$ and the performance remained same at stage 3 (refer back to Figure \ref{fig:alpha} (left)). At stage $4$, with $90\%$ supervision from teacher, we get an improved mAP of $0.383$ (overall 8.8\% improvement on base WELS-Net). 

\noindent \textbf{Only Teacher Networks ($\alpha = 0$) in SeCoST}: 
In Figure \ref{fig:alpha} (left), we showed that training the student using only the teacher's output ($1-\alpha = 1$) as the target already gives a better model. We get $0.366$ mAP versus $0.352$ when training only on ground truth labels. A similar observations has been made in \cite{hinton2015distilling}.
This may be because the soft probability outputs from the teacher network provides richer information compared to the `hard' ground truth labels. However, as shown in Figure \ref{fig:alpha_var} (right), this teacher-alone strategy does not work well for future generations. Here we run SeCoST for $3$ stages using only teacher's supervision. We see that the first stage leads to an improvement over the base network, and then there is saturation. This shows that co-supervision is necessary. Hence, the teacher`s knowledge needs to be coupled with the ground truth labels while training the sequence of students. 

\begin{figure}[t]
\centering
\includegraphics[width=0.7\linewidth]{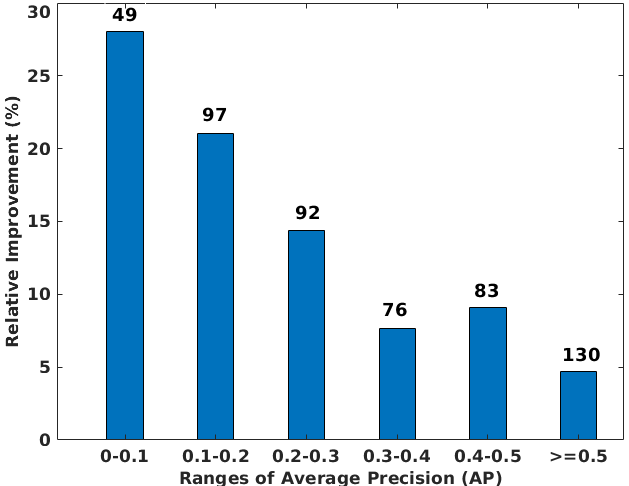}
\caption{\textbf{Left}: Relative Improvement in mean AP within different ranges. For instance, $0-0.1$ represents sound classes for which AP using base WELS-Net is $<0.1$. $y$-axis shows relative improvement in mean AP. Number at top of each bar shows number of classes with AP in that range. }
\label{fig:ap_range}
\end{figure}

\noindent \textbf{Some Class Specific Analysis}: 
We observe that SeCoST improves performance for $>85\%$ percent of the sound classes in Audioset ($448$ out of $527$). Of these, for $110$ classes we get $>20\%$ improvement in AP. Specifically, for \emph{Crushing, Harmonic and Mouse} sounds in Audioset vocabulary, we observe more than $100\%$ improvement using SeCoST. On the other hand, there are only 12 classes with more than 10\% drop in performance. Sound class \emph{Squish} has maximum drop in performance, around 19\%. Note that, these summaries are based on the best SeCoST model with mAP of $0.383$ and base WELS-Net with 0.352 mAP.
To further analyze class specific performance, we try to see whether the improvements are coming for classes where the base model already does well or if the classes with low APs are actually improving. Figure \ref{fig:ap_range} shows this relative mAP improvement for classes with APs within a specified range. For classes with the APs $<0.1$, the mean AP improves by $\sim28\%$. For classes with APs $\in[0.1,0.2)$, the gain is $21\%$. This show that SeCoST leads to considerable improvement in classes which are harder to learn for WELS-Net. For relatively easy classes (WELS-Net AP $>0.5$), we see $\sim4.5\%$ improvement. 

\section{Conclusion}
\label{sec:conc}
We proposed a sequential co-supervision learning framework for audio event detection. Our proposal, SeCoST, builds a generation of networks by designing student models that learn to predict a convex combination of teachers' predictions instead of the given ground truth. We showed that SeCoST gives a considerably better performance on Audioset compared to baseline model and state of the art performace. 
We note that the proposed framework is generally applicable to learning from noisy, weak labels, and we intend to future investigate the theoretical merits of the model in the future.


\bibliographystyle{IEEEbib}
\bibliography{references}

\end{document}